# Discrimination, artificial intelligence, and algorithmic decision-making

**Study by Prof. Frederik Zuiderveen Borgesius**
**Professor of Law, Institute for Computing and**
**Information Sciences (iCIS), Radboud University**
**Nijmegen, and Researcher at the Institute for Information**
**Law, University of Amsterdam (the Netherlands)**

*The opinions expressed in this work are the responsibility of the author and do not necessarily reflect the official policy of the Council of Europe.*

# TABLE OF CONTENTS

## EXECUTIVE SUMMARY

This report, written for the Anti-discrimination department of the Council of Europe, concerns discrimination caused by algorithmic decision-making and other types of artificial intelligence (AI). AI advances important goals, such as efficiency, health and economic growth but it can also have discriminatory effects, for instance when AI systems learn from biased human decisions.

In the public and the private sector, organisations can take AI-driven decisions with far-reaching effects for people. Public sector bodies can use AI for predictive policing for example, or for making decisions on eligibility for pension payments, housing assistance or unemployment benefits. In the private sector, AI can be used to select job applicants, and banks can use AI to decide whether to grant individual consumers credit and set interest rates for them. Moreover, many small decisions, taken together, can have large effects. By way of illustration, AI-driven price discrimination could lead to certain groups in society consistently paying more.

The most relevant legal tools to mitigate the risks of AI-driven discrimination are non-discrimination law and data protection law. If effectively enforced, both these legal tools could help to fight illegal discrimination. Council of Europe member States, human rights monitoring bodies, such as the European Commission against Racism and Intolerance, and Equality Bodies should aim for better enforcement of current non-discrimination norms.

But AI also opens the way for new types of unfair differentiation (some might say discrimination) that escape current laws. Most non-discrimination statutes apply only to discrimination on the basis of protected characteristics, such as skin colour. Such statutes do not apply if an AI system invents new classes, which do not correlate with protected characteristics, to differentiate between people. Such differentiation could still be unfair, however, for instance when it reinforces social inequality.

We probably need additional regulation to protect fairness and human rights in the area of AI. But regulating AI in general is not the right approach, as the use of AI systems is too varied for one set of rules. In different sectors, different values are at stake, and different problems arise. Therefore, sector-specific rules should be considered. More research and debate are needed.

## I.    INTRODUCTION

This report, written for the Anti-discrimination department of the Council of Europe, concerns risks of discrimination caused by algorithmic decision-making and other types of artificial intelligence (AI).

AI advances important goals, such as efficiency, health and economic growth. Our society relies on AI for many things, including spam filtering, traffic planning, logistics management, speech recognition, and diagnosing diseases. AI and algorithmic decision-making may appear to be rational, neutral and unbiased but, unfortunately, AI and algorithmic decision-making can also lead to unfair and illegal discrimination. As requested, the report focuses on the following questions.

1. *In which fields do algorithmic decision-making and other types of AI create discriminatory effects, or could create them in the foreseeable future?*
2. *What regulatory safeguards (including redress mechanisms) regarding AI currently exist, and which safeguards are currently being considered?*
3. *What recommendations can be made about mitigating the risks of discriminatory AI, to organisations using AI, to Equality Bodies in Council of Europe member states, and to human rights monitoring bodies, such as the European Commission against Racism and Intolerance?*
4. *Which types of action (legal, regulatory, self-regulatory) can reduce risks?*

This report uses the word "discrimination" to refer to objectionable or illegal discrimination, for instance on the basis of gender, skin colour, or racial origin.[1] The report speaks of "differentiation" when referring to discrimination in a neutral, unobjectionable, sense.[2]

This report focuses on only one risk in relation to algorithmic decision-making and AI: the risk of discrimination. Many AI-related topics are thus outside the scope of this report, such as automated weapon systems, self-driving cars, filter bubbles, singularity, data-driven monopolies, the risk that AI or robots cause mass unemployment. Also out of scope are privacy-related questions regarding the massive amounts of personal data that are collected to power AI-systems.

The report relies on literature review. Because of length constraints, this report should be seen as a quick scan, rather than an in-depth mapping of all relevant aspects of AI, algorithmic decision-making, and discrimination. I would like to thank Bodó Balázs, Janneke Gerards, Dick Houtzager, Margot Kaminski, Dariusz Kloza, Gianclaudio Malgieri, Stefan Kulk, Linnet Taylor, Michael Veale, Sandra Wachter and Bendert Zevenbergen for their valuable suggestions.

The remainder of the report is structured as follows. Chapter II introduces artificial intelligence, algorithmic decision-making, and some other key phrases. Next, the report discusses the above-mentioned questions. Chapter III maps fields where AI leads or might lead to discrimination. Chapter IV discusses regulatory safeguards. Chapter V highlights how organisations can prevent discrimination when using AI. The chapter also offers recommendations to Equality Bodies and human rights monitoring bodies on mitigating the risks of discriminatory AI and algorithmic decision-making. Chapter VI gives suggestions on improving regulation, and chapter VII provides concluding thoughts.

---

[1] In line with legal tradition, I use the words "racial origin" and "race" in this report. However, I do not accept theories that claim that there are separate human races.

[2] The purpose of algorithmic decision-making is often to discriminate (in the sense of differentiate or distinguish) between individuals or entities. See in detail the different meanings of "discrimination": Lippert-Rasmussen 2014.

## II. ARTIFICIAL INTELLIGENCE AND ALGORITHMIC DECISION-MAKING

The phrases AI and algorithmic decision-making are used in various ways, and there is no consensus about definitions. Below artificial intelligence (AI), algorithmic decision-making and some related concepts are briefly introduced.

### Algorithm

An algorithm can be described as "an abstract, formalised description of a computational procedure."[3] In this report, "decision" simply refers to the output, finding, or outcome of that procedure. As a rough rule of thumb, one could think of an algorithm as a computer program.

Sometimes, an algorithm decides in a fully automatic fashion. For instance, a spam filter for an e-mail service can filter out, fully automatically, spam messages from the user's inbox. Sometimes, humans make decisions assisted by algorithms; such decisions are *partly* automatic. For example, based on an assessment of a customer's credit by an AI system, a bank employee may decide whether a customer can borrow money from the bank.

However, when discussing discrimination, many risks are similar for fully and partly automated decisions. Recommendations by computers may have an air of rationality or infallibility, and people might blindly follow them. As Wagner et al. note, "the human being may often be led to "rubber stamp" an algorithmically prepared decision, not having the time, context or skills to make an adequate decision in the individual case."[4] Human decision-makers may also try to minimise their own responsibility by following the computer's advice.[5] The tendency to believe computers or to follow their advice is sometimes called "automation bias".[6] (We see in section IV.2 that some legal rules do distinguish fully and partly automated decisions.[7])

### Artificial intelligence

Artificial intelligence (AI) is, loosely speaking, "the science of making machines smart".[8] More formally, AI concerns "the study of the design of intelligent agents."[9] In this context, an agent is "something that acts", such as a computer.[10]

AI is a broad research field, which exists since the 1940s.[11] There are many types of AI. For instance, in the 1970s and 1980s, there was much research into "expert systems", "programs for reconstructing the expertise and reasoning capabilities of qualified specialists within limited domains."[12] Researchers programmed computers to answer questions, using preformulated answers. Such expert systems had some commercial success in the 1980s.[13] Expert systems had two disadvantages, observes Alpaydin. First, the logical rules in the systems did not always fit the messy reality of the world. "In real life, things are not true or false, but have grades of truth: a person is not either old or not old, but oldness increases gradually with age."[14] Second, experts had to provide the knowledge (the answers) to put into the systems. That process costs a lot of time and money.[15]

---

[3] Dourish 2016, p. 3. See also Domingos 2015.

[4] Wagner et al. 2018, p. 8. See also Broeders et al. 2017, p. 24-25.

[5] Zarsky 2018, p. 12.

[6] Parasuraman and Manzey 2010. See also Citron 2007, p. 1271-1272; Rieke, Bogen and Robinson 2018, p. 11.

[7] See the discussion of Article 22 GDPR in that section.

[8] Royal Society 2017, p. 16.

[9] Russel and Norvig 2016, p. 2, citing Poole, Mackworth and Goebel 1998, p. 1: "Computational Intelligence is the study of the design of intelligent agents."

[10] Russel and Norvig 2016, p. 4.

[11] Two early publications are: Turing 1951 and McCarthy et al. 1955.

[12] Puppe 1993, p. 3.

[13] Alpaydin 2016, p. 51.

### Machine learning

In the past decade, one type of AI has been particularly successful: machine learning.[16] With machine learning, the knowledge in the system does not have to be provided by experts. "In contrast, machine learning systems are set a task and given a large amount of data to use as examples of how this task can be achieved or from which to detect patterns. The system then learns how best to achieve the desired output."[17]

As a rough rule of thumb, machine learning could be summarised as "data-driven predictions".[18] Lerh and Ohm give a more detailed description: "machine learning refers to an automated process of discovering correlations (sometimes alternatively referred to as relationships or patterns) between variables in a dataset, often to make predictions or estimates of some outcome."[19]

Machine learning has become widely used during the past decade, in part because more and more data have become available to train the machines. Machine learning is so successful that nowadays many people say AI when they refer to machine learning (which is a type of AI).[20]

Related phrases are data mining, big data and profiling. Data mining, a type of machine learning, is "the process of discovering interesting patterns from massive amounts of data."[21] Data mining is also referred to as "knowledge discovery from data".[22] The phrase "big data" roughly refers to analysing large data sets.[23] "Profiling" involves automated data processing to develop profiles that can be used to make decisions about people.[24]

### Terminology in this report

Regarding technology, this report sacrifices precision for readability, and uses "AI", "AI system", "AI decision" etc, without specifying whether AI refers to machine learning or another technology. Thus, in this report, an "AI system" can refer, for instance, to a computer running an algorithm that was fed data by its human operators.

For ease of reading, this report uses phrases such as "effects of AI", almost as if AI is an entity that acts on its own. However, AI systems do not spontaneously come into existence. As Wagner et al. note, "Mathematic or computational constructs do not by themselves have adverse human rights impacts but their implementation and application to human interaction does."[25] Indeed, when an AI system makes decisions, it was an organisation that decided to use AI for that task.

In practice, an organisation that starts using AI rarely makes all relevant decisions about the AI system itself. An organisation might deploy an AI system, for which many

---

important choices have been made already.[26] In some cases, the effects of certain decisions in a pre-procurement or design stage of an AI system may only become apparent when the system is deployed in the real world. Apart from that, organisations can consist of many people, such as managers, lawyers and IT specialists. Nevertheless, for brevity, the report sometimes says that "organisations" do things. The next chapter discusses how AI can lead to discrimination and highlights areas where AI leads or might lead to discriminatory effects.

### III.    DISCRIMINATION RISKS

*In which fields do algorithmic decision-making and other types of AI create discriminatory effects, or could create them in the foreseeable future?*

#### 1.    HOW AI CAN LEAD TO DISCRIMINATION

This section discusses how AI can lead to discrimination; the next section gives examples where AI has led, or might lead, to discrimination. AI systems are often "black boxes".[27] It is often unclear for somebody why a system makes a certain decision about him or her. Because of the opaqueness of such decisions, it is difficult for people to assess whether they were discriminated against on the basis of, for instance, racial origin.

AI-driven decision-making can lead to discrimination in several ways. In a seminal paper, Barocas and Selbst distinguish five ways in which AI decision-making can lead, unintentionally, to discrimination.[28] The problems relate to (i) how the "target variable" and the "class labels" are defined; (ii) labelling the training data; (iii) collecting the training data; (iv) feature selection; and (v) proxies. In addition, (vi), AI systems can be used, on purpose, for discriminatory ends.[29] We discuss each problem in turn.

#### 1)    Defining the "target variable" and "class labels"

AI involves computers that find correlations in data sets. For instance, when a company develops a spam filter, the company feeds the computer e-mail messages that are labelled by humans as "spam" and "non-spam". Those labelled messages are the training data. The computer finds which characteristics of e-mail messages correlate with being labelled as spam. The set of discovered correlations is often called "model" or "predictive model". For instance, messages that are labelled as spam might often contain certain phrases ("magic weight loss pill", "millions of dollars for you" etc), or might be sent from certain IP addresses. As Barocas and Selbst put it, "by exposing so-called 'machine learning' algorithms to examples of the cases of interest (previously identified instances of fraud, spam, default, and poor health), the algorithm "learns" which related attributes or activities can serve as potential proxies for those qualities or outcomes of interest."[30] Such an outcome of interest is called a "target variable".

"While the target variable defines what data miners are looking for", explain Barocas and Selbst, "'class labels' divide all possible values of the target variable into mutually exclusive categories."[31] For spam filtering, people roughly agree about the class labels: which messages are spam or not.[32] But for some situations, it is less obvious what the target variable should be. "Sometimes," note Barocas and Selbst, "defining the target

---

variable involves the creation of *new* classes."[33] Suppose a company wants an AI system to sort job applications to find good employees. How is a "good" employee to be defined? In other words: what should be the "class labels"? Is a good employee one who sells the most products? Or one who is never late at work?

Some target variables and class labels, explain Barocas and Selbst, "may have a greater or lesser adverse impact on protected classes."[34] Suppose, for instance, that poorer people rarely live in the city centre and must travel further to their work than other employees. Therefore, poorer people are late for work more often than others because of traffic jams or problems with public transport. The company could choose "rarely being late often" as a class label to assess whether an employee is "good". But if people with an immigrant background are, on average, poorer and live further from their work, that choice of class label would put people with an immigrant background at a disadvantage, even if they outperform other employees in other aspects.[35] In sum, discrimination can creep into an AI system because of how an organisation defines the target variables and class labels.

### 2)    The training data: labelling examples

AI decision-making can also have discriminatory results if the system "learns" from discriminatory training data. Barocas and Selbst describe two ways in which biased training data can have discriminatory effects. First, the AI system might be trained on biased data. Second, problems may arise when the AI system learns from a biased sample.[36] In both cases, the AI system will reproduce that bias.

The training data can be biased because they represent discriminatory human decisions. Such a situation occurred at a medical school in the UK in the 1980s.[37] The school received many more applications than it could place. Therefore, the school developed a computer program to help sort the applications. The training data for the computer program were the admission files from earlier years, when people selected which applicants could enter medical school. The training data showed the computer program which characteristics (the input) correlated with the desired output (being admitted to the medical school). And the computer reproduced that selection system.

It turned out that the computer program discriminated against women and against people with an immigrant background. Apparently, in the years that provided the training data, the people that selected the students were biased against women and people with an immigrant background. As the British medical journal noted, "the program was not introducing new bias but merely reflecting that already in the system."[38] In sum, if the training data are biased, the AI system risks reproducing that bias.

### 3)    Training data: data collection

The sampling procedure can also be biased. For instance, when collecting data about crime, it could be the case that the police stopped more people with an immigrant background in the past. As Lum and Isaac note, "If police focus attention on certain ethnic groups and certain neighbourhoods, it is likely that police records will systematically over-represent those groups and neighbourhoods."[39]

---

[33] Barocas and Selbst 2016, p. 679.

[34] Barocas and Selbst 2016, p. 680.

[35] See Peck 2013.

[36] Barocas and Selbst 2016, p. 680-681.

[37] Lowry and Macpherson 1988; Barocas and Selbst 2016, p. 682.

[38] Lowry and Macpherson 1988.

[39] Lum and Isaac 2016, p. 15.

If an AI system is trained on such a biased sample, it will learn that people with an immigrant background are more likely to commit crime. Lum and Isaac note: "if biased data is used to train these predictive models, the models will reproduce (…) those same biases."[40]

The effects of such a biased sample could even be amplified by AI predictions. Suppose the police pay extra attention in a neighbourhood with many immigrants, while that neighbourhood has average crime levels. The police register more crime in that neighbourhood than elsewhere. Because the numbers show more crime is registered (and thus seems to occur) in that neighbourhood, even more policemen are sent there. This way, policing on the basis of crime statistics can cause a feedback loop.[41]

To give another example: poor people may be under-represented in a data set. This can be illustrated with Street Bump, a smartphone application that uses features such as GPS feeds to report road conditions to the city council. The Street Bump site explains: "Volunteers use the Street Bump mobile app to collect road condition data while they drive. The data provides governments with real-time information to fix problems and plan long-term investments."[42] If there are fewer smartphone users among poor people than among wealthier people, poor people are likely to be undercounted. The effect could be that faulty roads in poor neighbourhoods are under-represented in the dataset and therefore receive fewer reparations. The Street Bump app was used in the city of Boston, and that city aims to correct for such bias in data collection.[43] But the example illustrates how data collection could inadvertently lead to a biased data set. To sum up: biased training data can lead to biased AI systems.

### 4) *Feature selection*

A fourth problem relates to the features (categories of data) that an organisation selects for its AI system. If an organisation wants to use AI to predict something automatically, it needs to simplify the world to be able to capture it in data.[44] As Barocas and Selbst note, an organisation must "make choices about what attributes they observe and subsequently fold into their analyses."[45]

Suppose that an organisation wants to predict automatically which job applicants will be good employees. It is not possible, or at least too costly, for an AI system to assess each job applicant completely. An organisation could focus, for instance, on certain features, or characteristics, of each job applicant.

By selecting certain features, the organisation might introduce bias against certain groups. For example, many employers in the US look for people who studied at famous and expensive universities. But it might be relatively rare for certain racial groups to study at those expensive universities. Therefore, it may have discriminatory effects if an employer selects job applicants on the basis of whether they studied at a famous university.[46] In sum, organisations can cause discriminatory effects by selecting the features that an AI system uses for prediction.

---

[40] Lum and Isaac 2016, p. 15.

[41] Lum and Isaac 2016, p. 16. See also Ferguson 2017; Harcourt 2008; Robinson and Koepke 2016.

[42] http://www.streetbump.org accessed 10 September 2018.

[43] Crawford 2013. See also Barocas and Selbst 2016, p. 685; Federal Trade Commission 2016, p. 27.

[44] Barocas and Selbst 2016, p. 688.

[45] Barocas and Selbst 2016, p. 688.

[46] Barocas and Selbst 2016, p. 689.

### 5)     Proxies

Another problem concerns proxies. Some data that are included in the training set may correlate with protected characteristics. As Barocas and Selbst point out, sometimes "criteria that are genuinely relevant in making rational and well-informed decisions also happen to serve as reliable proxies for class membership."[47]

Suppose that a bank uses an AI system, trained on data covering the last twenty years, to predict which loan applicants will have problems repaying the loan. The training data do not contain information about protected characteristics such as skin colour. The AI system learns that people from postal code F-67075 were likely to default on their loans and uses that correlation to predict defaulting. Hence, the system uses what is at first glance a neutral criterion (postcode) to predict defaulting on loans. But suppose that the postcode correlates with racial origin. In that case, if the bank acted on the basis of this prediction and denied loans to the people in that postcode, the practice would harm people from a certain racial origin.

Barocas and Selbst explain that "[t]he problem stems from what researchers call "redundant encodings", cases in which membership in a protected class happens to be encoded in other data. This occurs when a particular piece of data or certain values for that piece of data are highly correlated with membership in specific protected classes."[48]

To illustrate: a dataset that does not contain explicit data about people's sexual orientation can still give information about people's sexual orientation. "Facebook friendships expose sexual orientation", found a study from 2009. The study "demonstrates a method for accurately predicting the sexual orientation of Facebook users by analysing friendship associations (…). [T]he percentage of a given user's friends who self–identify as gay male is strongly correlated with the sexual orientation of that user."[49]

The proxy problem is difficult to solve. Barocas and Selbst note: "Computer scientists have been unsure how to deal with redundant encodings in datasets. Simply withholding these variables from the data mining exercise often removes criteria that hold demonstrable and justifiable relevance to the decision at hand."[50] Hence, "[t]he only way to ensure that decisions do not systematically disadvantage members of protected classes is to reduce the overall accuracy of all determinations."[51]

### 6)     Intentional discrimination

Another situation can also occur: discrimination on purpose.[52] For example, an organisation could intentionally use proxies to discriminate on the basis of racial origin. As Kroll et al. observe: "A prejudiced decisionmaker could skew the training data or pick proxies for protected classes with the intent of generating discriminatory results".[53] When an organisation uses proxies, the discrimination would be harder to detect than when the organisation openly discriminates.

To give a hypothetical example: an organisation could discriminate against pregnant women, while that discrimination would be difficult to discover. The US retail store Target reportedly constructed a "pregnancy prediction" score, based on around 25 products, by analysing the shopping behaviour of customers. If a woman buys some

---

of those products, Target can predict with reasonable accuracy that she is pregnant. Target wanted to reach people with advertising during moments in life when they are more likely to change their shopping habits. Therefore, Target wanted to know when female customers were going to give birth. "We knew that if we could identify them in their second trimester, there's a good chance we could capture them for years".[54] Target used the prediction for targeted marketing, but an organisation could also use such a prediction for discrimination.[55]

To sum up, AI decision-making can lead to discrimination in at least six ways, which relate to (i) the definition of the target variables and the class labels; (ii) the labelling and (iii) collecting of the training data; (iv) the selection of the features; (v) proxies. And (vi) organisations could use AI systems to discriminate on purpose. AI can also lead to other types of unfair differentiation, or to errors. We return to those topics in chapter VI.

## 2. FIELDS IN WHICH AI BRINGS DISCRIMINATION RISKS

This section provides examples of fields where AI decision-making has led, or could lead, to discrimination.

### Police, crime prevention

We start with the public sector. A notorious example of an AI system with discriminatory effects is the system known as "Correctional Offender Management Profiling for Alternative Sanctions" – COMPAS for short.[56] The COMPAS system is used in parts of the US to predict whether defendants will commit crime again. The idea is that COMPAS can help judges to determine whether somebody should be allowed to go on probation (supervision outside prison). The COMPAS system does not use racial origin or skin colour as an input. But research by Angwin et al., investigative journalists at ProPublica, showed in 2016 that COMPAS is "biased against blacks."[57] ProPublica summarises:

> COMPAS (…) correctly predicts recidivism 61 percent of the time. But blacks are almost twice as likely as whites to be labelled a higher risk but not actually reoffend. It makes the opposite mistake among whites: They are much more likely than blacks to be labelled lower risk but go on to commit other crimes.[58]

Moreover, "Black defendants were also twice as likely as white defendants to be misclassified as being a higher risk of violent recidivism. And white violent recidivists were 63 percent more likely to have been misclassified as a low risk of violent recidivism, compared with black violent recidivists."[59]

Northpointe, the company behind COMPAS, disputes that the system is unfair.[60] ProPublica and Northpointe disagree mainly on what standard of fairness should be used to assess the system.[61] Academic statisticians have argued that, in some cases, different standards of fairness are incompatible mathematically, which has consequences for what discrimination prevention should or could look like. ProPublica was concerned about what can be called "disparate mistreatment", where different

---

groups receive different error types disproportionately (for instance individuals from some groups having a higher possibility of being deemed high-risk when they would not go on to commit a crime). Yet another important characteristic of risk scores is that they are correctly "calibrated". This means that for a group of individuals deemed to have an 80% chance of going on to commit a crime, 80% of that group indeed do go on to commit a crime. This should also be the same within groups, such as within black or white defendants. If this were not the case, then judges would need to interpret "high risk" for a black defendant differently than the same "high risk" for a white defendant, which brings other biases into play. Statisticians have indicated that where the underlying propensity to recidivism does differ, it is mathematically impossible to also have equalised error rates.[62]

Sometimes the police use AI systems for predictive policing: automated predictions about who will commit crime, or when and where crime will occur.[63] As noted above, predictive policing systems can reproduce and even amplify existing discrimination.

### Selection of employees and students

In the private sector, AI can have discriminatory effects as well. We saw, for instance, that AI can be used to select prospective employees or students. As the example of the medical school in the UK showed, an AI system could lead to discrimination because of biased training data. Reportedly, Amazon stopped using an AI system for screening job applicants because the system was biased against women. In the words of Reuters, "the company realised its new system was not rating candidates for software developer jobs and other technical posts in a gender-neutral way."[64] Based on historical training data, "Amazon's system taught itself that male candidates were preferable."[65]

### Advertising

AI is used for targeted online advertising, a very profitable sector for some companies (Facebook and Google, both among the world's most valuable companies, derive most of their profit from online advertising[66]). Online advertising can have discriminatory effects. Sweeney showed in 2013 that, when people searched for African-American-sounding names, Google displayed advertisements that suggested that somebody had an arrest record. For white-sounding names, Google displayed fewer ads suggestive of arrest records. Presumably, Google's AI system analysed people's surfing behaviour and inherited a racial bias.[67]

Datta, Tschantz, and Datta simulated identical internet users who self-declared as male or female in settings. The researchers then analysed the ads that Google presented.[68] "Google showed the simulated males ads from a certain career coaching agency that promised large salaries more frequently than the simulated females, a finding suggestive of discrimination."[69] Researchers note that it is unclear why women were shown fewer ads for high-paying jobs, because of the opaqueness of the system: "We cannot determine who caused these findings due to our limited visibility into the ad ecosystem, which includes Google, advertisers, websites and users."[70]

---

This is an example where the opaqueness of AI systems makes it harder to discover discrimination and its cause. People could be discriminated against without being aware. If an AI system targets job ads only at men, women might not realise that they are excluded from the ad campaign.[71]

The Dutch Data Protection Authority found that Facebook enabled advertisers to target people based on sensitive characteristics. For instance, "data relating to sexual preferences were used to show targeted advertisements".[72] The Data Protection Authority says that Facebook amended its practices to make such targeting impossible.[73] Angwin and Perris, at ProPublica, showed that "Facebook lets advertisers exclude users by race. Facebook's system allows advertisers to exclude black, Hispanic and other "ethnic affinities" from seeing ads."[74] ProPublica also showed that some firms use Facebook's targeting possibilities to advertise job ads only to people under a certain age.[75] Spanish researchers showed that "Facebook labels 73% of EU users with sensitive interests", such as "Islam", "reproductive health", and "homosexuality".[76] Advertisers can target advertising on the basis of such interests.

### Price discrimination

Online shops can differentiate the price for identical products based on information the shop has about a consumer: a practice called online price differentiation. A shop can recognise website visitors, for instance through cookies, and categorise them as price-sensitive or price-insensitive. With price differentiation, shops aim to charge each consumer the maximum price that he or she is willing to pay.[77]

Princeton Review, a US company that offers online tutoring services, charged different prices in different areas in the US, ranging from 6600 to 8400 dollars. Presumably, the costs for delivering the service were the same for each area, as the company offers its tutoring service over the Internet. Angwin et al. found that the company's price differentiation practice led to higher prices for people with an Asian background: "Customers in areas with a high density of Asian residents were 1.8 times as likely to be offered higher prices, regardless of income."[78] The company probably did not set out to discriminate on the basis of racial origin. Perhaps the company had tested different prices in different neighbourhoods and found that in certain areas people bought the same amount of services, even for higher prices. Nevertheless, the effect was that certain ethnic groups paid more.

### Image search and analysis

Systems to search for images can also have discriminatory effects. In 2016, a search in Google Images for "three black teenagers" led to mugshots, while a search for "three white kids" mostly lead to pictures of happy white kids. In response to shocked reactions, Google said: "Our image search results are a reflection of content from across the web, including the frequency with which types of images appear and the way they're described online. (…) This means that sometimes unpleasant portrayals of sensitive subject matter online can affect what image search results appear for a given query."[79] Indeed, one could say that Google's AI system merely reflected society.[80] But even if the fault lies with society rather than with the AI system, those image search results could influence people's beliefs.

---

[71] Munoz, Smith and Patil, 2016, p. 9; Zuiderveen Borgesius 2015a, chapter 3, section 3.

[72] Dutch Data Protection Authority 2017; Dutch Data Protection Authority 2017a.

[73] Dutch Data Protection Authority 2017.

[74] Angwin and Perris 2016. See also Angwin, Tobin and Varner 2017. Dalenberg 2017 examines the application of EU non-discrimination law to ad targeting. In 2018, NGOs filed a lawsuit in the USA against Facebook for discrimination under US fair housing laws, for allowing the exclusion of women, disabled veterans and single mothers from a housing advertisement's potential audience (Bagli 2018).

[75] Angwin, Scheiber and Tobin 2017.

[76] Cabañas, Cuevas and Cuevas 2018. Such interests are defined as "special categories" of data, also called "sensitive data", in European data protection law. See article 9 of the European Union's General Data Protection Regulation. See, on data protection law: section IV.2.

[77] Zuiderveen Borgesius and Poort 2017.

[78] Angwin, Mattu and Larson 2015; Larson, Mattu and Angwin 2015.

Kay, Matuszek and Munson found that "image search results for occupations slightly exaggerate gender stereotypes and portray the minority gender for an occupation less professionally. There is also a slight under-representation of women."[81]

A different type of problem concerns image recognition by AI systems. Some image recognition software has difficulties in recognising and analysing non-white faces. Facial-tracking software by Hewlett Packard did not recognise dark-coloured faces as faces.[82] And the Google Photos app labelled a picture of an African-American couple as "gorillas".[83] A Nikon camera kept asking people from an Asian background: "Did someone blink?"[84] An Asian man had his passport picture rejected, automatically, because "subject's eyes are closed" – but his eyes were open.[85] Buolamwini and Gebru found that "darker-skinned females are the most misclassified group (with error rates of up to 34.7%). The maximum error rate for lighter-skinned males is 0.8%."[86] Perhaps some of the errors mentioned above were the result of only training systems on pictures of white men.

### Translation tools

The AI behind automated translation tools can also reflect inequality and discrimination. If people type "He is a doctor. She is a nurse" into Google Translate and translate the phrases into Turkish, Google Translate provides: "O bir hemşire. O bir doktor". Those Turkish sentences are gender-neutral; Turkish does not differentiate between the words "he" and "she". When translating the Turkish text into English again, Google Translate provides: "She is a nurse. He is a doctor".

The example is taken from research by Caliskan, Bryson and Narayanan, which shows "that machines can learn word associations from written texts and that these associations mirror those learned by humans."[87] In other words, "natural language necessarily contains human biases, and the paradigm of training machine learning on language corpora means that AI will inevitably imbibe these biases as well."[88]

Prates, Avelar and Lamb tested twelve gender-neutral languages, such as Hungarian and Chinese, in Google Translate. The authors wrote sentences such as "he/she is an engineer" in the gender-neutral languages and translated the sentences into English with Google Translate. The authors concluded that Google Translate "exhibits a strong tendency towards male defaults".[89] Moreover, "male defaults are not only prominent but exaggerated in fields suggested to be troubled with gender stereotypes, such as STEM (Science, Technology, Engineering and Mathematics) jobs."[90] In sum, AI-driven translation tools can provide results that reflect existing gender inequality. Perhaps such results could also worsen inequality, as they could influence people's ideas.

---

[79] Google's reaction, quoted in York 2016.

[80] Allen 2016.

[81] Kay, Matuszek and Munson 2015.

[82] Frucci 2009.

[83] BBC News 2015. See also Noble 2018.

[84] Sharp 2009.

[85] Regan 2016.

[86] Buolamwini and Gebru 2018.

[87] Caliskan, Bryson and Narayanan 2017.

[88] Narayanan 2016.

[89] Prates, Avelar and Lamb 2018, p. 1.

[90] Prates, Avelar and Lamb 2018, p. 28.

### Nuancing the risks

We saw that AI decision-making could have discriminatory effects – but AI systems do not necessarily perform worse than humans. Unfortunately, many humans also make discriminatory decisions. Indeed, in some cases, AI systems discriminate because they were trained on data that reflect discrimination by humans. Hence, it makes a difference whether one compares AI decision-making with human decisions in the real world (which, unfortunately, are sometimes discriminatory) or with hypothetical decisions in an ideal world without discrimination.[91] Of course, the goal should be a world without any unfair or illegal discrimination.

Apart from that, AI could also be used to discover discrimination or inequality.[92] Suppose an AI system shows that a collection of stock photos contains gender stereotypes. One way of interpreting such a finding is that the AI system illustrates stereotyped behaviour that already exists. Hence, an AI system could help to discover existing inequality that might have remained hidden otherwise.

## IV.     LEGAL AND REGULATORY SAFEGUARDS

***What regulatory safeguards (including redress mechanisms) regarding AI currently exist, and which safeguards are currently being considered?***

Non-discrimination law and data protection law are the main legal regimes that could protect people against AI-driven discrimination. This chapter discusses each regime in turn and highlights other potentially relevant fields of law and self-regulation. The chapter paints with a broad brush and focuses on the core principles of legal regimes. Issues lying outside the scope of this report include differences in regulation in Council of Europe member States, the territorial scope of laws and enforcement of laws against organisations in other States.

### 1. NON-DISCRIMINATION LAW

Discrimination is prohibited in many treaties and constitutions, including the European Convention on Human Rights.[93] Article 14 of the European Convention on Human Rights states:

> "The enjoyment of the rights and freedoms set forth in this Convention shall be secured without discrimination on any ground such as sex, race, colour, language, religion, political or other opinion, national or social origin, association with a national minority, property, birth or other status.[94]"

Both *direct* and *indirect* discrimination are prohibited by the European Convention on Human Rights.[95] Direct discrimination means, roughly summarised, that people are discriminated against on the basis of a protected characteristic, such as racial origin. The European Court of Human Rights describes direct discrimination as follows: "there must be a difference in the treatment of persons in analogous, or relevantly similar, situations", which is based "on an identifiable characteristic".[96] EU law non-discrimination law uses a similar definition.[97]

---

[91] See also Tene and Polonetsky 2017.

[92] See Munoz, Smith and Patil 2016, p. 14.

[93] See e.g. Article 7 of the United Nations Declaration of Human Rights; Article 26 of the International Covenant on Civil and Political Rights; Article 21 of the Charter of Fundamental Rights of the European Union.

[94] Protocol 12 to that Convention lays down a similar prohibition with, regarding certain aspects, a broader scope. "*The enjoyment of any right set forth by law* shall be secured without discrimination on any ground such as sex, race, colour, language, religion, political or other opinion, national or social origin, association with a national minority, property, birth or other status." Article 1, Protocol No. 12 to the Convention for the Protection of Human Rights and Fundamental Freedoms, European Treaty Series - No. 177, Rome, 4.XI.2000. On 18 September 2018, the total number of ratifications of/accessions to Protocol 12 stood at 20. See, for an up-to-date list: https://www.coe.int/en/web/conventions/search-on-treaties/-/conventions/treaty/177/signatures?p_auth=0Kq9rtcm.

[95] While the European Convention on Human Rights has some horizontal effect, the Convention does not directly regulate discrimination in the private sector.

Indirect discrimination occurs, roughly speaking, when a practice is neutral at first glance but ends up discriminating against people of a certain racial origin (or another protected characteristic).[98] Indirect discrimination is called "disparate impact" in the United States. Indirect discrimination is described as follows by the European Court of Human Rights:

> "[A] difference in treatment may take the form of disproportionately prejudicial effects of a general policy or measure which, though couched in neutral terms, discriminates against a group. Such a situation may amount to "indirect discrimination", which does not necessarily require a discriminatory intent."[99]

Indirect discrimination is defined similarly in EU law:

> "Indirect discrimination shall be taken to occur where an apparently neutral provision, criterion or practice would put persons of a racial or ethnic origin at a particular disadvantage compared with other persons, unless that provision, criterion or practice is objectively justified by a legitimate aim and the means of achieving that aim are appropriate and necessary."[100]

AI decision-making can unintentionally lead to indirect discrimination. Regarding indirect discrimination, the law focuses on the effects of a practice, rather than on the intention of the alleged discriminator.[101] Hence, it is not relevant whether the discriminator had the intention to discriminate.

Non-discrimination law can be used to fight discriminatory AI decisions. For instance, AI decisions that make people from a certain racial background pay more for goods and services could breach the prohibition of indirect discrimination. With AI decision-making, accidental indirect discrimination probably occurs more often than intentional discrimination.

However, non-discrimination law has several weaknesses in the context of AI decision-making. The prohibition of indirect discrimination does not provide a clear and easily applicable rule.[102] The concept of indirect discrimination results in rather open-ended standards, which are often difficult to apply in practice. It needs to be proven that a seemingly neutral rule, practice or decision disproportionately affects a protected group and is thereby *prima facie* discriminatory. In many cases, statistical evidence is used to show such a disproportionate effect.[103]

---

The European Court of Human Rights accepts that such a suspicion of indirect discrimination can be rebutted if the alleged discriminator can invoke an objective justification:

> "A general policy or measure that has disproportionately prejudicial effects on a particular group may be considered discriminatory even where it is not specifically aimed at that group and there is no discriminatory intent. This is only the case, however, if such policy or measure has no "objective and reasonable" justification".[104]

Such a justification must be objective and reasonable, and a measure, practice or rule does not meet these requirements if it:

> "has no objective and reasonable justification, that is if it does not pursue a legitimate aim or if there is not a reasonable relationship of proportionality between the means employed and the aim sought to be achieved".[105]

Along similar lines, EU law says that a practice will *not* constitute indirect discrimination if it "is objectively justified by a legitimate aim and the means of achieving that aim are appropriate and necessary".[106] Whether an alleged discriminator can invoke such an objective justification depends on all the circumstances of a case and requires a nuanced proportionality test.[107] Therefore, it is not always clear whether a certain practice breaches the prohibition of indirect discrimination.

The requirement that a *prima facie* case of indirect discrimination must be shown may also cause difficulties, since this type of discrimination can remain hidden. Suppose that somebody applies for a loan on the website of a bank. The bank uses an AI system to decide on such requests. If the bank automatically denies a loan to a customer on its website, the customer does not see why the loan was denied. Moreover, the customer cannot see whether the bank's AI system denies loans to a disproportionate percentage of, for instance, women.[108] So even if customers knew that an AI system rather than a bank employee decided, it would be difficult for them to discover whether the AI system is discriminatory.

Another weakness relates to non-discrimination law's concept of protected characteristics. Non-discrimination statutes typically focus on (direct and indirect) discrimination based on protected characteristics, such as race, gender or sexual orientation.[109] But many new types of AI-driven differentiation seem unfair and problematic – some might say discriminatory – while they remain outside the scope of most non-discrimination statutes. Hence, non-discrimination law leaves gaps. In section IV.3, we return to such unfair types of differentiation that might escape non-discrimination law.

In conclusion, non-discrimination law, in particular through the concept of indirect discrimination, prohibits many discriminatory effects of AI. However, enforcement is difficult, and non-discrimination law has weaknesses. The next section takes a look at data protection law.

---

## 2. DATA PROTECTION LAW

Data protection law is a legal tool that aims to defend fairness and fundamental rights, such as the right to privacy and the right to non-discrimination.[110] Data protection law grants rights to people whose data are being processed (data subjects)[111] and imposes obligations on parties that process personal data (data controllers).[112] Eight principles form the core of data protection law; they can be summarised as follows:

(a) Personal data may only be processed lawfully, fairly and transparently ("lawfulness, fairness, and transparency").

(b) Such data may only be collected for a purpose that is specified in advance, and should not be used for other unrelated purposes ("purpose limitation").

(c) Such data should be limited to what is necessary for the processing purpose ("data minimisation").

(d) Such data should be sufficiently accurate and up-to-date ("accuracy").

(e) Such data should not be retained for an unreasonably long period ("storage limitation").

(f) Such data should be secured against data breaches, illegal use etc ("integrity and confidentiality").[113]

(g) The data controller is responsible for compliance ("accountability").[114]

These principles are included in the Council of Europe's Data Protection Convention 108 (revised in 2018[115]) and the European Union's General Data Protection regulation (GDPR, from 2016). Similar principles are included in more than a hundred national data privacy laws in the world.[116]

Data protection law could help mitigate risks of unfair and illegal discrimination.[117] For instance, data protection law requires transparency about personal data processing. Therefore, organisations must provide information, for instance in a privacy notice, about all stages of an AI decision-making process that involve personal data.[118] It is true that most people do not read privacy notices.[119] Nevertheless, such notices could be helpful for researchers, journalists, and supervisory authorities. If a privacy notice suggests that a processing practice could have discriminatory effects, authorities can investigate.

Under certain circumstances, the GDPR and Data Protection Convention 108 require organisations (data controllers) to conduct a data protection impact assessment (DPIA). An impact assessment can be described as follows:

---

[110] See Article 1(2) and recital 71, 75, and 85 GDPR, and Article 1 of the COE Data Protection Convention 2018; Council of Europe Big Data Guidelines 2017, article 2.3.

[111] Article 4(1) GDPR; Article 2(a) COE Data Protection Convention 2018.

[112] Article 4(7) GDPR; Article 2(d) COE Data Protection Convention 2018.

[113] Article 5(1)(a)-5(1)(f) GDPR; Articles 5, 7, and 10 COE Data Protection Convention 2018.

[114] Article 5(2) of the GDPR; Article 10(1) COE Data Protection Convention 2018.

[115] Article 5, 7, and 10 COE Data Protection Convention 2018.

[116] Greenleaf 2017.

[117] See, on the interplay between data protection law and discrimination law: Schreurs et al. 2008; Gellert et al. 2013; Hacker 2018; Lammerant, De Hert, Blok 2017.

[118] Article 5(1)(a); Article 13; Article 14 GDPR; Articles 5(4)(a) and 8 COE Data Protection Convention 2018.

[119] Zuiderveen Borgesius 2015.

An impact assessment is a tool used for the analysis of possible consequences of an initiative on a relevant societal concern or concerns, if this initiative can present dangers to these concerns, with a view to supporting informed decision-making whether to deploy this initiative and under what conditions, ultimately constituting a means to protect these concerns.[120]

The GDPR requires a DPIA when a practice is "likely to result in a high risk to the rights and freedoms of natural persons", especially when using new technologies.[121] In some circumstances, the GDPR always requires a DPIA (because the GDPR assumes a high risk), for instance when organisations take fully automated decisions that have legal or similar effects for people.[122] Hence, for many AI systems that make decisions about people, the GDPR requires a DPIA.[123] The risk of unfair or illegal discrimination must also be considered when conducting a DPIA.[124]

Under the Council of Europe's Data Protection Convention 108, and under the Charter of Fundamental Rights of the European Union, each member State must have an independent Data Protection Authority.[125] Such Data Protection Authorities must have powers of investigation.[126] The GDPR gives most details about the investigative powers of Data Protection Authorities. A Data Protection Authority can, for instance, obtain access to premises of controllers, carry out investigations in the form of data protection audits and order data controllers to provide information and to give access to their data processing systems.[127]

### Rules on automated decisions

The GDPR contains specific rules for certain types of "automated individual decision-making".[128] These rules aim, among other things, to mitigate the risk of illegal discrimination.[129] The Council of Europe's Data Protection Convention also contains rules on automated decisions, which are less detailed than in the GDPR.[130] Here, we focus on the GDPR.

Article 22 of the GDPR, sometimes called the Kafka provision, contains an in-principle prohibition of fully automated decisions with legal or similar significant effects and applies, for instance, to fully automated e-recruiting practices without human intervention.[131] The predecessor of the GDPR already had a similar provision, which has not been applied much in practice.[132] The main rule of the GDPR's provision on automated individual decision-making reads as follows:

---

[120] Kloza et al. 2017, p. 1. See also Article 29 Working Party 2017 (WP248); Binns 2017; Mantelero 2017; Wright and De Hert 2012.

[121] Article 25(1) GDPR.

[122] Article 35(3)(a) GDPR. See also recital 91 GDPR.

[123] Article 35(3)(b) and 35 (3)(c) GDPR could also apply to some AI systems.

[124] Article 29 Working Party 2017 (WP248), p. 6, p. 14. See also Kaminski 2018a, p. 25; Edwards and Veale 2017.

[125] Article 8(3) of the Charter of Fundamental Rights of the European Union. See also Article 51 GDPR; chapter IV COE Data Protection Convention.

[126] Chapter VI GDPR; chapter IV COE Data Protection Convention.

[127] Article 58(1) GDPR. The Data Protection Authority can also exercise these rights against "processors", organisations that process personal data for data controllers.

[128] Article 22 GDPR. The discussion of the GDPR's rules on automated decisions is based on and includes sentences from Zuiderveen Borgesius and Poort 2017.

[129] See Recital 71 GDPR.

[130] Article 9(1)(a) COE Data Protection Convention 2018.

[131] Recital 71 GDPR.

[132] Korff 2012. The predecessor was Article 15 of the Data Protection Directive. That Article 15 was based on a provision of the Data Protection Act of France from 1978. See Bygrave 2001.

> The data subject shall have the right not to be subject to a decision based solely on automated processing, including profiling,[133] which produces legal effects concerning him or her or similarly significantly affects him or her.[134]

Roughly summarised: people may not be subjected to certain automated decisions with far-reaching effects. The GDPR says people have a "right not to be subject to" certain decisions. But it is generally assumed that this right implies an in-principle prohibition of such decisions.[135]

Slightly rephrasing Mendoza and Bygrave, four conditions must be met for the provision to apply: (i) there is a decision, which is based (ii) solely (iii) on automated data processing; (iv) the decision has legal or similarly significant effects for the person.[136]

An example of a decision with "legal effects" would be a court decision, or decision regarding a social benefit granted by law, such as pension payments.[137] An example of a decision with "similarly significantly" effects would be a bank that denies credit automatically.[138] And Data Protection Authorities say that online price differentiation could "similarly significantly affect" somebody, if it leads to "prohibitively high prices [that] effectively bar someone from certain goods or services."[139]

There are exceptions to the in-principle prohibition of certain automated decisions. In short, the prohibition does not apply if the automated decision (i) is based on the individual's explicit consent; (ii) is necessary for a contract between the individual and the data controller; or (iii) is authorised by law.[140]

If a controller can rely on the (i) consent or (ii) contract exception to bypass the prohibition, a different rule is triggered: "the data controller shall implement suitable measures to safeguard the data subject's rights and freedoms and legitimate interests, at least the right to obtain human intervention on the part of the controller, to express his or her point of view and to contest the decision".[141] Hence, in some circumstances, the data subject can ask for a human to reconsider the automated decision. For instance, a bank could ensure that customers can call the bank to have a human reconsider the decision, if the bank automatically denies them a loan through the bank's website.

In addition to its general transparency requirements, the GDPR also contains transparency requirements specific to automated decisions:

> [T]he controller shall provide the data subject with the following information (...) the existence of automated decision-making, including profiling (...) and, at least in those cases, meaningful information about the logic involved, as well as the significance

---

[133] The GDPR defines "profiling" as follows: "'Profiling' means any form of automated processing of personal data consisting of the use of personal data to evaluate certain personal aspects relating to a natural person, in particular to analyse or predict aspects concerning that natural person's performance at work, economic situation, health, personal preferences, interests, reliability, behaviour, location or movements." Art. 4(4) GDPR.

[134] Art. 22 GDPR.

[135] De Hert and Gutwirth 2008; Korff 2012; Wachter, Mittelstadt, and Floridi 2017; Zuiderveen Borgesius 2015a.

[136] Mendoza and Bygrave 2017.

[137] See Article 29 Working Party 2018 (WP251), p. 21.

[138] Recital 71 GDPR. See for more examples that "could" constitute automated decisions that "similarly significantly affect" people: Article 29 Working Party 2018 (WP251), p. 22.

[139] Article 29 Working Party 2018 (WP251). p. 22.

[140] Article 29 Working Party 2018 (WP251), p. 22.

[141] Article 22(3) GDPR. As Kaminski notes, the GDPR's text "creates a version of algorithmic due process: a right to an opportunity to be heard." Kaminski 2018a, p. 8.

and the envisaged consequences of such processing for the data subject.[142]

Hence, in some cases, an organisation would have to explain that it uses AI decision-making and would have to provide meaningful information about the logic of that process.

There has been a great deal of scholarly attention as to whether the GDPR's rules on automated decisions create a "right to explanation" of individual decisions.[143] Recital 71 suggests the existence of an individual right to "explanation" of AI decisions – a right that could be useful to protect fairness.[144]

Many scholars are sceptical of whether such a right would be effective, noting for instance that many types of automated decisions remain outside the scope of the GDPR's rules.[145] To illustrate: the GDPR's automated decision provision only applies to decisions based "solely" on automated processing. Hence, when a bank employee denies a loan on the basis of a recommendation by an AI system, as long as the employee is not rubber-stamping, the provision does not apply.[146]

It remains to be seen what the practical effect of these GDPR provisions will be. As noted, the predecessor of the GDPR provision on automated decisions has remained a dead letter. Regardless, the attention to the GDPR provisions has helped to foster an interdisciplinary discussion on explaining AI decisions.

The modernised Convention 108 appears more generous for individuals in its phrasing around explanation rights. Unlike the GDPR provision, which applies to decisions that have significant effect and are "solely" based on automated processing, Convention 108 gives individuals a right "to obtain, on request, knowledge of the reasoning underlying data processing where the results of such processing are applied to him or her".[147] The breadth of what it means to "apply" a "result" is yet to be seen in practice in any national implementations.

*Caveats*

Several caveats are in order regarding data protection law's possibilities as a tool to fight AI-driven discrimination. First, there is a compliance and enforcement deficit. Data Protection Authorities have limited resources. And many Data Protection Authorities do not have the power to impose serious sanctions (in the EU, such authorities received new powers with the GDPR). Previously, many organisations did not take compliance with data protection law seriously.[148] It appears that compliance improved with the arrival of the GDPR, but it is too early to tell.

Second, parts of algorithmic processes are outside the scope of data protection law. Data protection law only applies when personal data are processed. It does not apply to predictive models because they do not relate to identifiable persons. For example, a predictive model that says "80% of the people living in postal code F-67075 pay their

---

[142] Article 13(2)(f) and 14(2)(f) GDPR.

[143] See for instance Edwards and Veale 2017; Goodman and Flaxman 2016; Kaminski 2018; Kaminski 2018a; Malgieri G and Comandé 2017; Mendoza and Bygrave 2017; Selbst and Powles 2017; Wachter et al. 2017.

[144] Recital 71: "such processing should be subject to suitable safeguards, which should include (..) the right (…) to obtain an explanation of the decision reached after such assessment".

[145] See for instance Edwards and Veale 2017; Wachter et al. 2017; Zuiderveen Borgesius 2015a, chapter 9, section 6.

[146] See Article 29 Working Party 2018 (WP251). The Working Party says that a superficial check by a human (rubber stamping) is not sufficient. However, as noted by Veale and Edwards 2018, it is not clear how organisations are supposed to ensure that decisions have non-superficial human input.

[147] Article 9(1)(c) COE Data Protection Convention 2018: "Every individual shall have a right (…) to obtain, on request, knowledge of the reasoning underlying data processing where the results of such processing are applied to him or her". See Veale and Edwards 2018.

[148] See Zuiderveen Borgesius 2015a, chapter 8, section 2.

bills late" is not a personal datum, as the model does not refer to an individual. (When a predictive model is applied to an individual, data protection law applies again.[149])

Third, data protection law uses many open and abstract norms, rather than black-and-white rules.[150] Data protection law must use open norms, because its provisions apply in many different situations, in the private and the public sector. This regulatory approach, an omnibus approach, has many advantages. For instance, the open norms do not have to be adapted each time when a new technology is developed. But one disadvantage is that the open norms can be difficult to apply.[151]

Fourth, data protection law has strict rules on "special categories" of data (sometimes called "sensitive data"), such as data regarding racial origin or revealing health status.[152] Those rules create challenges for assessing and mitigating discrimination. Many of the methods to tackle discrimination in AI systems implicitly assume that organisations hold these sensitive data – yet to meet data protection law, many organisations may not be holding them. Tension remains between respecting data protection law and collecting sensitive data to fight discrimination.[153]

Fifth, even where explanations of AI decisions might be legally required by the GDPR or Convention 108, it is often difficult to explain the logic behind a decision, when an AI system, analysing large amounts of data, arrives at that decision.[154] And in some cases, it is not clear how much an explanation would help people, especially insofar as it places the burden on them to understand the decision and its appropriateness.[155]

That said, more transparency and explanation of AI decisions could be useful. For more than a decade, scholars have been calling for the development of transparency-enhancing technologies (TETs), to enable meaningful transparency regarding automated decision-making.[156] Such technologies should "aim at making information flows more transparent through feedback and awareness thus enabling individuals as well as collectives to better understand how information is collected, aggregated, analysed and used for decision-making."[157] Computer scientists are exploring various forms of explainable AI.[158]

In any case, it is much too early to assess the effect of the modernised Convention 108 and the GDPR. More legal research is needed on how data protection law could help to mitigate discrimination risks.[159] While data protection law is largely untested as a non-discrimination tool, it does offer possibilities to fight illegal discrimination.

---

[149] See Zuiderveen Borgesius 2015a, chapter 2 and chapter 5. See, on the weaknesses of data protection law in the area of AI decision-making: Wachter and Mittelstadt 2018.

[150] Zuiderveen Borgesius 2015a, chapter 9, section 1.

[151] See, on different types of legal rules: Chapter VI, section 1.

[152] Article 9 GDPR; article 6 COE Data Protection Convention 2018. The strict rules for special categories of data aim, in part, to fight discrimination. See on "special categories of data" in the context of AI: Malgieri and Comandé 2017a.

[153] Goodman 2016; Ringelheim and De Schutter 2008; Ringelheim and De Schutter 2009; Veale and Binns 2017; Žliobaitė and Custers 2016. Some methods to audit AI systems while maintaining privacy using cryptography are emerging. See Kilbertus et al. 2018.

[154] Ananny and Crawford 2016; Burrell 2016; Binns et al. 2018; Edwards and Veale 2017; Hildebrand 2015; Kroll et al. 2016; 2018; Wachter, Mittelstadt and Russell 2017.

[155] Edwards and Veale 2017.

[156] Hildebrandt and Gutwirth 2008, chapter 17.

[157] Diaz and Gürses 2012.

[158] See Guidotti et al. 2018; Miller 2017; Selbst and Barocas 2018; Tickle et al. 1998. See also this Google project: "What If... you could inspect a machine learning model, with no coding required?", https://pair-code.github.io/what-if-tool/index.html#about accessed 1 October 2018. That project took inspiration from Wachter, Mittelstadt and Russell 2017.

[159] Researchers are starting to explore how data protection law can help to fight discrimination. See for instance: Goodman 2016; Mantelero 2018; Hacker 2018; Hoboken and Kostic (forthcoming); Wachter 2018; Wachter and Mittelstadt 2018.

### 3. OTHER REGULATION

In the area of AI decisions, other fields of law could also help to ensure fairness, and perhaps help to mitigate discrimination-related problems. For example, consumer law could be invoked to protect consumers against some types of manipulative AI-driven advertising.[160] As discriminatory behaviour by a company causes more problems when the company has a monopoly position, competition law could also help to protect people.[161] For the public sector, administrative law and criminal law could be relevant to protect fair procedures.[162] Freedom of information laws could be used to obtain information about public sector AI systems.[163] But the application of these fields of law to protect people in the area of AI is largely unexplored. A discussion of those fields of law falls outside the scope of this report.

***Regulation under consideration***

Several regulatory measures that could be relevant for AI-driven discrimination are currently being considered. The Council of Europe's Consultative Committee of the Convention for the Protection of Individuals with regard to Automatic Processing of Personal Data published a draft report in September 2018: "Artificial intelligence and data protection: challenges and possible remedies."[164]

The Council of Europe's Steering Committee on Media and Information Society has set up an expert committee on AI: the Committee of Experts on human rights dimensions of automated data processing and different forms of artificial intelligence. The expert committee will conduct studies and give guidance for possible future standard-setting.[165]

The European Union is active in the area of AI too. In 2018, the European Commission published a communication on AI, and has set up a High-Level Expert Group on Artificial Intelligence,[166] which is tasked with proposing draft AI Ethics Guidelines.[167] The EU Agency for Fundamental Rights is also examining AI.[168] Furthermore, in 2017 the Commission proposed an ePrivacy Regulation to protect privacy on the Internet, which could be relevant for AI and machine learning, as it would limit the collection of certain types of privacy-sensitive data on the Internet.[169]

An EU Regulation from 2016 concerns one type of AI decision: algorithmic trading on stock exchanges etc. The Regulation states: "An investment firm shall ensure that its compliance staff has at least a general understanding of how the algorithmic trading systems and trading algorithms of the investment firm operate."[170] Moreover, "an investment firm shall establish and monitor its trading systems and trading algorithms

---

through a clear and formalised governance arrangement".[171] Perhaps similar requirements could be adopted for other sectors.

### Self-regulation

Several organisations have proposed principles that aim for fair, accountable or ethical AI. For example, the organisation FATML, Fairness, Accountability, and Transparency in Machine Learning, published "Principles for accountable algorithms and a social impact statement for algorithms"[172] The principles call for organisations to "ensure that algorithmic decisions do not create discriminatory or unjust impacts when comparing across different demographics (eg race, sex, etc)."[173]

There are other self-regulatory principles on ethics and AI, often less focused on discrimination. Examples include the Asilomar AI principles of the (US-based) Future of Life Institute,[174] the Montreal Declaration for Responsible AI[175] and the Principles for ethical AI of the UNI Global Union.[176] IEEE, a technical professional organisation, launched a Global Initiative on Ethics of Autonomous and Intelligent Systems.[177] A "Partnership on AI to Benefit People and Society" was set up by Apple, Amazon, DeepMind and Google, Facebook, IBM and Microsoft, to study and formulate best practices on AI technologies.[178] In principle, such self-regulation principles are laudable. Ethical AI is obviously better than unethical AI. Self-regulatory principles could help to mitigate discrimination problems and could provide inspiration for law-makers.

However, protecting human rights cannot be left to self-regulation or soft law.[179] The main problem is that self-regulation is non-binding. Moreover, the above-mentioned principles are often somewhat abstract and do not give detailed guidance.[180] Wagner warns against "ethics washing" in the context of AI: "much of the debate about ethics seems increasingly focussed on private companies avoiding regulation. Unable or unwilling to properly provide regulatory solutions, ethics is seen as the "easy" or "soft" option which can help structure and give meaning to existing self-regulatory initiatives."[181] Indeed, self-regulation and soft law should not distract from a possible need for (hard) legal regulation. Chapter VI discusses how the law could be improved. But first we turn to recommendations to organisations using AI, and to human rights monitoring bodies and Equality Bodies.

---

[171] Article 1, idem.

[172] https://www.fatml.org/resources/principles-for-accountable-algorithms accessed 24 September 2018.

[173] https://www.fatml.org/resources/principles-for-accountable-algorithms accessed 24 September 2018.

[174] https://futureoflife.org/ai-principles/ accessed 24 September 2018.

[175] https://www.montrealdeclaration-responsibleai.com/the-declaration accessed 24 September 2018.

[176] http://www.thefutureworldofwork.org/opinions/10-principles-for-ethical-ai/ accessed 24 September 2018.

[177] https://standards.ieee.org/industry-connections/ec/autonomous-systems.html accessed 24 September 2018. See also Koene et al. 2018.

[178] https://www.partnershiponai.org/about/ accessed 24 September 2018. See for a list of, and critique of, other ethics principles for AI: Greene, Hoffman and Stark 2018.

[179] See, generally on self-regulation and fundamental rights: Angelopoulos et al. 2016.

[180] See Campolo et al. 2017, p. 34.

[181] Wagner 2018. See also Nemitz 2018.

## V. RECOMMENDATIONS

***What recommendations can be made on mitigating the risks of discriminatory AI, to organisations using AI, to Equality Bodies in Council of Europe member States, and to human rights monitoring bodies, such as the European Commission against Racism and Intolerance?***

### 1. ORGANISATIONS USING AI

Several measures are important for public and private organisations wishing to prevent discrimination when they use AI. Such measures include education, obtaining technical and legal expertise, and careful planning of AI projects.

### *Education*

Education is important to make organisations realise the risks of accidental AI-driven discrimination. Relevant employees of an organisation – including managers, lawyers, and computer scientists – should be aware of the risks. As we have seen, in many examples of discriminatory AI, the organisations did not set out to discriminate. If such organisations had been aware of the risks, they might have been able to prevent that discrimination. Perhaps education could also help to mitigate the effects of "automation bias" among employees.[182]

### *Risk assessment and mitigation*

When an organisation starts an AI project, it should perform risk assessment and risk mitigation. This entails (i) involving individuals from multiple disciplines, such as computer science and law, to define the risks of a project; (ii) recording both the assessment and mitigation processes; (iii) monitoring the implementation of a project; and (iv) often reporting outward in some way, either to the public or to an oversight body.[183]

Organisations should ensure that they receive help from computer scientists who understand discrimination risks. (The phrase "computer scientist" is used here as shorthand. Data scientists or other and people with sufficient knowledge of AI could also provide expertise). An emerging field in computer science focuses on discrimination risks in the field of AI decisions. Since 2014, an organisation called FATML organises workshops and conferences, with the aim of "[b]ringing together a growing community of researchers and practitioners concerned with fairness, accountability and transparency in machine learning."[184] Computer scientists have published promising results, for instance on discrimination-aware data mining.[185]

Defining the risks of an AI project can be challenging. When left alone, computer scientists have to make value-laden decisions while building an AI system, and often find risks or choices hard to communicate to senior decision-makers.[186] Assessing and mitigating discrimination risks requires active support for those developing AI systems, and the time and money needed for this should be an active consideration in all relevant projects.

The risks and applicable legal and normative principles are different for each sector. Different risks are involved for an AI system that selects job applicants, for example, than for one that predicts crime. Therefore, experts with knowledge of a particular

---

sector should be involved.[187] It may be useful to set up an ethics committee to assess and discuss AI systems that entail risks for human rights.[188] It can also be useful to bring in academics, civil society groups and potentially impacted individuals to discuss their concerns over the system.[189]

One way to assess the risks of an AI project is to carry out an appropriate type of impact assessment. Inspiration can be drawn from the GDPR's DPIA requirement for certain risky data processing operations.[190] And organisations – especially in the public sector – should consider publishing the impact assessment report.

The risks of the AI system should also be monitored during its use, particularly as the phenomena the AI system is modelling are likely to change over time, and the risks and impacts may change with them.[191] Organisations should consider publishing yearly reports monitoring the system.

It is often possible to prevent, or at least minimise, discriminatory effects. For instance, an organisation can choose not to use certain features as input data in their AI system. To illustrate: one US company that helps to select employees says that it does not use "distance to work" as a factor to predict which applicants will be successful employees, because that factor correlates too much with race. As reported by The Atlantic: "The distance an employee lives from work, for instance, is never factored into the score given each applicant, although it is reported to some clients. That's because different neighbourhoods and towns can have different racial profiles, which means that scoring distance from work could violate equal-employment-opportunity standards."[192]

At AI companies and university research labs, the workforce is often not diverse – largely male and white for instance. Such organisations might pay more attention to discrimination when they have a more diverse workforce. Hence, organisations should aim to hire a more diverse workforce.[193] Obviously, aiming for a more diverse workforce is always important.

***Public sector bodies***

Compared to the private sector, the public sector has extra responsibilities. Indeed, many legal rules, for instance in the field of human rights, criminal procedure law and administrative law, aim to protect people against the powerful State. The extra responsibilities also apply when public sector bodies use AI systems.

Therefore, where possible, AI systems in the public sector should be designed for transparency.[194] In some situations, information about AI systems could be released to the public for scrutiny, in the spirit of the Open Data movement. Yet in some cases, such information might leak personal data and create privacy risks[195] or might allow people to game the AI system.[196] Therefore, public bodies might want to enable controlled access to their AI systems for researchers or civil society in secure environments, much as statistical agencies do to sensitive microdata today.[197]

---

Furthermore, public sector could adopt a sunset clause when introducing AI systems that take decisions about people. Such a sunset clause could require that a system should be evaluated, say after three years, to assess whether it brought what was hoped for.[198] If the results are disappointing, or if the disadvantages or the risks are too great, consideration should be given to abolishing the system. While public sector bodies have extra responsibilities, private sector organisations such as companies can take similar measures to those proposed above for the public sector.

## 2. EQUALITY BODIES AND HUMAN RIGHTS MONITORING BODIES

What recommendations can be made to Equality Bodies in Council of Europe member States and to human rights monitoring bodies, such as the European Commission against Racism and Intolerance, on mitigating the risks of AI-driven discrimination?

### *Education and technical expertise*

Equality Bodies and human rights monitoring bodies should be aware of the promises and threats of AI. Therefore, education for Equality Bodies and monitoring bodies on the basics of AI and its risks is needed.

Equality Bodies and human rights monitoring bodies should also ensure that they obtain technical expertise on AI, by involving computer scientists.[199] Computer scientists can recognise and understand certain risks better than, for instance, lawyers.[200] Computer scientists, even if they are not AI specialists, could carry out certain types of investigations into AI-driven discrimination. As Rieke, Bogen and Robinson note, "Scrutiny doesn't have to be sophisticated to be successful."[201] Problems with an AI system can often be discovered through "simple observation of a system's inputs and outputs".[202] And computer scientists who are not AI specialists themselves often know which specialists to hire for certain investigations. Depending on budget, Equality Bodies and human rights monitoring bodies could hire computer scientists for a project, or on a more permanent basis.

Equality Bodies and human rights monitoring bodies should consider organising public awareness campaigns for organisations in the public and private sector.[203] As noted, in many cases, organisations use discriminatory AI systems by accident. Awareness could help.

More generally, schools and universities that teach computer science, data science, AI, and related topics should teach students about human rights and ethics. Many universities already offer such courses to computer science students.[204] Equality Bodies and human rights monitoring bodies could consider assisting schools and universities with such courses.[205]

To permit public debate, it would be good if the general public knew more about the risks of discriminatory AI – and about the many advantages and possibilities of AI. However, awareness building should not lead to responsibilisation. This term describes "the process whereby subjects are rendered individually responsible for a task which previously would have been the duty of another – usually a state agency – or would not have been recognized as a responsibility at all."[206] Policy-makers should not make people responsible for defending themselves against discrimination.[207] That said, awareness is important for an inclusive debate on the risks of AI decisions.

---

[198] McCray, Oye and Petersen 2010; Broeders, Schrijvers and Hirsch Ballin, p. 23.

[199] As mentioned, the phrase "computer scientist" is used in this report as shorthand. Data scientists or other people with sufficient knowledge of AI could also provide expertise.

[200] See, on the importance of technical expertise for Data Protection Authorities: Raab and Szekely 2017.

[201] Rieke, Bogen and Robinson 2018, p. 2.

[202] Rieke, Bogen and Robinson 2018, p. 8. They also give examples of scrutiny of AI systems (p. 31-34).

[203] See ECRI Statute Resolution 2002, Article 12; ECRI general policy recommendation no. 2 (2018), para. 13(e); para. 34, and explanatory memorandum para. 64.

[204] Fiesler 2018 compiled a list of more than 200 courses on tech ethics.

[205] See ECRI general policy recommendation no. 10: on combating racism and racial discrimination in and through school education, 15 December 2006, Strasbourg, CRI(2007)6 https://rm.coe.int/ecri-general-policy-recommendation-no-10-on-combating-racism-and-racia/16808b5ad5 accessed 14 October 2018.

### Prior consultation with Equality Bodies

Equality Bodies could require public sector bodies to discuss with them any planned projects that involve AI decision-making about individuals or groups. For instance, an Equality Body could help to assess whether training data are biased.[208] Equality Bodies could also require each public sector body using AI decision-making about people to ensure that it has sufficient legal and technical expertise to assess and monitor risks. And public sector bodies could be required to regularly assess whether their AI systems have discriminatory effects. (Depending on the national situation, Equality Bodies could also suggest, rather than require).

Equality Bodies and human rights monitoring bodies could help to develop a specific method for a "human rights and AI impact assessment". As mentioned, impact assessments can be useful – but to date, there is no specific impact assessment method for AI.[209] When developing such a method, different stakeholders and people from different disciplines should be involved. Inspiration can be drawn from privacy and data protection impact assessments.[210]

### Engage in public procurement processes

Equality Bodies should seek, through national provisions and processes as well as through lobbying for increased access, to be involved in the procurement of public-sector AI systems from an early stage. Equality Bodies can help ensure that concerns around discrimination are built into the AI systems being procured: that systems are open enough to audit and subject to appropriate safeguards.

### Cooperate with Data Protection Authorities

As said, for AI-driven discrimination, the two most relevant legal frameworks are non-discrimination law and data protection law. It would be a shame if those fields of law operate in their own silos.[211] Equality Bodies should cooperate with Data Protection Authorities. For instance, it could be helpful to exchange knowledge and to learn from one another's experiences.[212] Many Data Protection Authorities have some technical expertise in house,[213] and some have experience with hiring outside computer scientists for research projects.[214] Data Protection Authorities may learn about organisations that use AI systems that entail discrimination risks, and could warn Equality Bodies. Equality Bodies could provide information to Data Protection Authorities, for instance about discrimination risks. Depending on the national situation, it could also be useful for Equality Bodies to cooperate with Consumer Protection Authorities and Competition law authorities.

---

For cooperation and knowledge sharing between different types of regulators, the European Data Protection Supervisor proposed in 2016 to set up "a voluntary network of regulatory bodies to share information (…) about possible abuses in the digital ecosystem and the most effective way of tackling them."[215] Perhaps that initiative could provide inspiration for Equality Bodies and human rights monitoring bodies.[216]

### Cooperate with academics

Equality Bodies and human rights monitoring bodies should keep in touch with, and perhaps cooperate with, academics. This report illustrates how many examples of discriminatory AI decisions were discovered by academic researchers (and by investigative journalists).[217] Many academics love to assist regulators but are not in regular contact with them. In the short term, Equality Bodies and monitoring bodies could visit conferences and other events where academic researchers meet. At many international privacy conferences, discriminatory AI is a much-debated topic. Several of these conferences attract a mix of regulators, practitioners, civil society groups and scholars from different disciplines, such as law, computer science, philosophy and sociology.[218] Equality Bodies and monitoring bodies could also consider organising conferences, round tables or other events on discrimination risks of AI, to foster contacts between the research community and Equality Bodies. And perhaps Equality Bodies and monitoring bodies could commission more research on AI's discrimination risks (see section VI.3) or set up a working party on AI's discrimination risks.[219]

Equality Bodies and human rights monitoring bodies should not only engage with civil society groups that work on discrimination[220] but also with consumer groups[221] and civil society groups that focus on technology policy and digital rights.[222] Civil society groups that work on discrimination often have different expertise from groups that work on technology and digital rights. More contact between such groups would be useful too, as many of them are interested in AI-driven discrimination.[223]

### Litigation and regulation

Depending on the national situation, Equality Bodies could also engage in strategic litigation in the area of AI decision-making.[224] And Equality Bodies and human rights monitoring bodies could push for regulation to mitigate discrimination risks of AI.[225] Suggestions to improve regulation are discussed in the next chapter.

---

[215] European Data Protection Supervisor 2016.

[216] As an aside: within universities too, more cooperation is needed between different types of legal scholars, such as non-discrimination law specialists (often working at human rights institutes) and data protection law specialists (often working at law and technology institutes).

[217] See Rieke, Bogen and Robinson 2018, p. 31.

[218] See for instance: the CPDP Computers, Privacy and Data Protection conference in Brussels https://www.cpdpconferences.org; the APC Amsterdam Privacy Conference https://www.apc2018.com; TILTing Perspectives https://www.tilburguniversity.edu/research/institutes-and-research-groups/tilt/events/tilting-perspectives; and the PLSC Privacy Law Scholars Conference http://law.berkeley.edu/plsc. The ACM Conference on Fairness, Accountability, and Transparency (ACM FAT*) will be in Amsterdam in 2020: https://www.fatml.org. All accessed 14 October 2018.

[219] See ECRI Statute Resolution 2002, Article 6(1); 6(2); ECRI general policy recommendation no. 2 (2018), article 13(d).

[220] See ECRI Statute Resolution 2002, Article 10(1) and 13.

[221] For consumer organisations, BEUC (the European Consumer Organisation) could be a point of contact. BEUC's members are 43 consumer organisations from 32 European countries. https://www.beuc.eu/about-beuc/who-we-are accessed 10 October 2018. See also European Consumer Organisation BEUC 2018.

[222] For groups focusing on rights and freedoms in the digital environment, European Digital Rights (EDRi) could be a point of contact. EDRi is an association of civil and human rights organisations from across Europe. https://edri.org/members/ accessed 10 October 2018.

[223] See Gangadharan and Niklas 2018, who interviewed NGOs and conclude that better cooperation is needed between (i) privacy- and technology-oriented NGOs and (ii) discrimination-oriented NGOs.

[224] See ECRI general policy recommendation no. 2 (2018), Article 14-16.

[225] See ECRI Statute Resolution 2002, Article 1; ECRI general policy recommendation no. 2 (2018), Article 13(j).

## VI. IMPROVING REGULATION

***Which types of action (legal, regulatory, self-regulatory) can reduce risks?***

Current law has weaknesses when applied to AI-driven discrimination, as we saw in chapter IV. Additional regulation is probably needed to protect people against illegal discrimination and unfair differentiation. Section 1 provides preliminary remarks about regulating in the area of fast-developing technology. Section 2 focuses on improving enforcement of existing non-discrimination norms. Section 3 discusses whether the legal norms themselves should be amended because of AI decision-making. The suggestions in this chapter are meant as starting points for discussion rather than as definitive policy advice.

### 1. REGULATION AND FAST-DEVELOPING TECHNOLOGY

Regulating brings extra challenges when the rules are to apply to fast-developing technology. Adopting statutes or treaties may take years or even decades. Meanwhile, technology, the market and society develop quickly.

These challenges are not unique for AI; there is experience with regulating new technologies. When regulating in the area of new technologies, policy-makers can combine different types of rules, such as statutes with broad principles and guidelines (by regulators for instance) with more specific rules.[226] The statutes could be phrased in a reasonably technology-neutral way. Technology-neutral legal provisions with broad principles have the advantage of not having to be changed every time a new technology is developed. A disadvantage is that broad principles can be difficult to apply in practice. Therefore, guidance by regulators can be useful.[227] Guidelines can be amended faster and can thus be more specific and concrete. Guidelines should be evaluated regularly and amended whenever required.[228]

Data protection law partly takes this combined approach.[229] Data protection law (such as the GDPR and the modernised Convention 108) contains many broadly phrased provisions that can be applied to different situations and technologies.[230] For instance, data protection law does not contain specific rules for CCTV, or for monitoring in the workplace. But as far as personal data (including on video images) are used, data protection law does apply to CCTV and workplace monitoring.

In addition to data protection law's statutory provisions, Data Protection Authorities often adopt interpretative guidelines with more specific and concrete requirements for different situations, such as CCTV,[231] the workplace[232] and automated decision-making.[233] In the EU, the European Data Protection Board and its predecessor have adopted more than 250 guidelines since 1995.[234] Similarly, the Council of Europe has

---

adopted guidelines in addition to the Data Protection Convention 108, for instance on big data,[235] the police sector[236] and profiling.[237]

Hence, if new legal rules were adopted to mitigate discrimination risks in the area of AI, perhaps statutory rules should be combined with a possibility for regulatory bodies to adopt guidelines that are easier to amend. There are more possibilities than statutory law and regulator guidance, such as co-regulation: self-regulation with varying degrees of influence of public regulators. The basic idea remains the same: different types of rules can be combined.[238] As Koops puts it, "Through multi-level legislation, open-ended formulations and a mixed approach of abstract and concrete rules that are periodically evaluated, adequate legal certainty with respect to current technologies may be ensured, while at the same time sufficient scope is given for future technological developments."[239]

Of course, there must be democratic legitimacy and sufficient checks and balances regarding entities that set rules or guidelines. In sum, regulating in the area of new technologies is hard, but possible – and often necessary.

## 2. ENFORCEMENT

### Improving enforcement of current non-discrimination norms

Regarding discrimination in the area of AI-driven decisions, the overarching norms are reasonably clear – in our society we do not, and should not, accept discrimination on the basis of protected characteristics such as racial origin. Below are some suggestions on enforcement of non-discrimination norms in the area of AI.

### Transparency

As noted, one of the problems with AI systems is the lack of transparency; their "black box" character.[240] The opaqueness can be seen as a problem in itself – but the opaqueness also makes it harder to discover discrimination.

Regulation can aim to improve transparency. The law (including guidelines etc) could, for instance, require that AI systems used in the public sector are developed in such a way that they enable auditing and explainability.[241] For the private sector too, such requirements could be considered.[242] There are precedents for such requirements in the private sector; a requirement of interpretability exists for certain systems for algorithmic trading.[243]

For some types of systems, it could be useful if public sector bodies release the underlying code (software). Sometimes, examining the code can provide information about how a system works. As Rieke, Bogen, and Robinson note, "code audits are most likely to be useful when there is a clearly defined question about how a software program operates in regulated space, and particular standards against which to measure a system's behaviour or performance."[244] Freedom of information laws could be adapted so that the code in AI systems is subject to such laws. Such an amendment would enable journalists, academics and others to obtain and examine such code.

---

AI systems are often protected by trade secrets, intellectual property rights or a company's terms and conditions.[245] Such protection makes it harder for regulators, journalists, and academics to investigate such systems. Perhaps the law should be adapted to improve research exceptions and to enable some types of research. And perhaps the law should require organisations to disclose certain information to researchers upon request. Such regulation must strike a delicate balance between public interest in transparency and commercial, privacy and other interests in opaqueness.[246]

In many cases, the code alone does not give much information about an AI system, as the system can only be assessed when it is used in practice. "For even moderately complex programs," observe Rieke, Bogen, and Robinson, "it may be necessary to see a program run "in the wild," with real users and data to truly understand its effects."[247]

The law could require the public sector to use only AI systems that have been properly assessed for risks and enable oversight and auditing.[248] A similar requirement could be considered for the private sector when AI systems are used for certain decisions, for instance on eligibility for insurance, credit or a job.[249] More research and debate is needed on who should conduct such audits. For oversight and auditing of AI systems, an organisation needs considerable expertise.[250]

***Investigation and enforcement powers***

Council of Europe member States should ensure that Equality Bodies and Data Protection Authorities receive adequate funding, and that they have sufficient investigation and enforcement powers.[251] Without enforcement, transparency will not necessarily lead to accountability.[252]

In sum, Equality Bodies and human rights monitoring bodies can push for regulation that enables better enforcement of current non-discrimination norms in the area of AI decision-making. However, AI decision-making also opens the way for new types of discrimination and differentiation that largely escape current non-discrimination and other laws. We turn to that topic now.

### 3. REGULATING NEW TYPES OF DIFFERENTIATION

Non-discrimination law and data protection law leave gaps in the context of AI.[253] Many non-discrimination statutes apply only to certain protected characteristics, such as race, gender or sexual orientation.[254] The statutes do not apply to discrimination on the basis of financial status for instance. Data protection law can help to fill some, but definitely not all, gaps in non-discrimination law.

AI systems can escape non-discrimination law when they differentiate on the basis of newly invented classes.[255] To give a simplified example: suppose an AI system finds a correlation between (i) using a certain web browser and (ii) a greater willingness to pay. An online shop could charge higher prices to people using that browser.[256] Such practices would remain outside the scope of non-discrimination law, as a browser type is not a protected characteristic. (For this hypothesis we assume that the browser type is not a proxy for a protected characteristic).

---

[245] See Bodo et al. 2017, p. 171-175; Malgieri 2016; Wachter and Mittelstadt 2018, p. 63-77.

[246] Similar questions arise in open data versus privacy discussions. See Zuiderveen Borgesius, Gray and Van Eechoud 2015.

[247] Rieke, Bogen and Robinson 2018, p. 19.

[248] See Campolo et al. 2018, p. 1.

[249] See Campolo et al. 2018, p. 1.

[250] It has been suggested that a specific oversight body for automated profiling (AI-driven decision-making) might be useful. See Koops 2008.

[251] See ECRI general policy recommendation no. 2 (2018), Article 28.

[252] See Kaminski 2018a, p. 21.

[253] See section IV.1 and IV.2.

[254] Gerards 2007; Khaitan 2015.

[255] Custers 2004. See also Mittelstadt et al. 2016.

### AI can reinforce social inequality

But AI decisions that remain outside the scope of non-discrimination law can still lead to differentiation that is unfair or has other drawbacks. For instance, insurance companies could use AI systems to set premiums for individual consumers, or to deny some consumers insurance. To some extent, risk differentiation is necessary, and an accepted practice, for insurance. And it can be considered fair when high-risk customers pay higher premiums.

But there are drawbacks. Too much risk differentiation could make insurance unaffordable for some consumers and could threaten the risk-pooling function of insurance. Furthermore, risk differentiation might result in the poor paying more. A consumer who lives in a poor neighbourhood with many burglaries might pay more for house insurance, because the risk of a burglary is higher. But if neighbourhoods where many poor people live have higher risks, then poor people pay, on average, more.[257]

More generally, AI could reinforce social inequality. For instance, Valentino-De Vries, Singer-Vine and Soltani showed that some online price differentiation practices in the US had the effect that people in poor areas paid higher prices. Several shops charged more to consumers who live in the countryside than to consumers in large cities.[258] In the countryside, consumers have to drive hours to visit a competitor. Therefore, an online shop does not have to use cheap prices; most customers will not drive for hours to buy the product at a cheaper price. In a large city, a consumer can easily go to a competitor to buy a product. Therefore, some online shops offered cheaper prices in large cities. This pricing scheme had the effect, probably unintentionally, that poorer people paid, on average, higher prices, as people tend to be poorer in the countryside of the US.[259] AI can thus reinforce social inequality. But, as noted, someone's financial status is not a protected characteristic, so non-discrimination law does not regulate such a practice (assuming that the practice does not lead to indirect discrimination based on a protected characteristic).[260]

### AI can lead to errors

Non-discrimination law has little to say about incorrect AI predictions (false positives and false negatives). A problem with AI decisions is that they are often incorrect for a particular individual. AI decision-making often entails applying a predictive model to individuals. A simplified example of a predictive model is: "80% of the people living in postal code F-67075 pay their bills late." If, based on this group profile, a company denies loans to all people in postal code F-67075, it also denies loans to the 20% who pay their bills on time.[261] Such practices could disproportionately harm certain groups

---

in society. Sometimes, an AI system makes more errors for minority groups than for the majority.[262]

### New rules?

Additional regulation should be considered, because AI decision-making that escapes non-discrimination law can still be unfair. But it is probably not useful to adopt rules for AI decision-making in general. AI is used in many different sectors and for many purposes, and often, AI does not threaten human rights.[263] An AI system of a chess computer does not bring the same risks as an AI system for predictive policing.

Even for AI systems that make decisions about humans, the risks are different in different sectors, and different rules should apply. The fairness of AI decision-making cannot be assessed in the abstract. In each sector, or application area, different arguments have different weights.[264] And in different sectors, different normative and legal principles apply. For instance, the right to a fair trial and the presumption of innocence are important in the field of criminal law. In consumer transactions, freedom of contract is an important principle. Hence, when new rules are considered, such rules need to focus on specific sectors.

Whether there is a need for new rules could be assessed as follows. For a particular sector, several questions should be answered.

(i) Which rules apply in this sector, and what are the rationales for those rules? A rule may, for example, aim to protect a human right, or express a legal principle, such as equality, contractual freedom, or the right to a fair trial. Economic rationales also differ from sector to sector. For instance, risk pooling is important for insurance, while it is not relevant in most other sectors. Hence, for each sector the rationales behind the rules differ.

(ii) How is or could AI decision-making be used in this sector, and what are the risks? For instance, false positives are a serious problem in the context of criminal law. A false positive could lead to people being questioned, arrested or perhaps even punished. We should not accept AI decision-making that breaches the underlying values of criminal law. By contrast: if an incorrect decision by an AI system for price discrimination makes a consumer pay extra, the effect is often less harmful than when an incorrect AI decision leads to someone being arrested by the police.

(iii) Considering the rationales for the rules in this sector, should the law be improved in the light of AI decision-making? Does AI threaten the law's underlying principles or undermine the law's goals? If current law leaves important risks unaddressed, amendments should be considered.

In conclusion, new rules may be needed for AI decision-making, to protect fairness and human rights such as the right to non-discrimination. However, more research and debate are required on the questions of whether and which rules are needed.

### Empirical and technical research

Information is necessary for good policy. There is a clear need for more information about AI-driven discrimination, and hence for more research.[265] Council of Europe member States should support research – research by human rights monitoring bodies, Equality Bodies, and by academics. More empirical research is needed for instance. It is unclear on what scale AI decision-making is used. How often does

---

algorithmic decision-making lead to discrimination (on the basis of racial origin for instance)? And to other types of unfair differentiation?

More computer science research into solutions is needed too. For instance, how could AI systems be designed so they respect and promote human rights, fairness and accountability? Can training data be checked for discrimination risks?[266] As noted, an emerging and vibrant field of computer science focuses on such questions.[267] More generally: if countries fund AI research, part of that funding should be used for research into the risks for fairness and human rights, and into mitigating those risks.

***Normative and legal research***

There is also a need for public debate, and for normative and legal research. How could the prohibition of indirect discrimination be enforced more effectively? How should the law deal with unfair differentiation that remains outside the scope of non-discrimination law? How to define fairness in diverse sectors? How should the law (and technology) protect people against intersectional[268] and structural discrimination?[269] Should the law protect some types of "group privacy", and how?[270] How to safeguard the rule of law when AI systems make decisions about people?[271] Which types of decisions, if any, should never be taken by computers? How could data protection law be used in practice to fight discrimination? Are new rules needed, or are tweaks to non-discrimination law and data protection law sufficient? Which tweaks would be needed? Which new rules would be needed?

## VII.    CONCLUSION

In conclusion, AI offers many exciting possibilities to improve our societies. But AI decision-making also brings risks – it is often opaque and can have discriminatory effects, for instance when an AI system learns from data reflecting biased human decisions.

In the public and the private sector, organisations can take AI-driven decisions with far-reaching effects for people. Public sector bodies can use AI for predictive policing or sentencing recommendations, and for decisions on, for instance, pensions, housing assistance or unemployment benefits. The private sector can also take AI decisions with major consequences for people, such as decisions regarding employment, housing or credit. Moreover, many small decisions, taken together, can have large effects. One targeted advertisement is rarely a major problem, but when aggregated, targeted advertising may exclude some groups. And AI-driven price differentiation could lead to certain groups in society consistently paying more.

The most relevant legal instruments to mitigate the risks of AI-driven discrimination are non-discrimination law and data protection law. If effectively enforced, both legal instruments could help to fight illegal discrimination. Council of Europe member States, human rights monitoring bodies, such as the European Commission against Racism and Intolerance, and Equality Bodies should aim for better enforcement of current non-discrimination norms.

---

But AI also paves the way for new types of unfair differentiation (or discrimination) that escape current laws. Most non-discrimination statutes only apply to discrimination on the basis of protected characteristics, such as racial origin. Such statutes do not apply if organisations differentiate on the basis of newly invented classes that do not correlate with protected characteristics. Such differentiation could still be unfair, however, for instance when it reinforces social inequality. We probably need additional regulation to protect fairness and human rights in the area of AI. But regulating AI in general is not the right approach, as the use of AI systems is too varied for one set of rules. We need sector-specific rules, because different values are at stake, and different problems arise, in different sectors. More debate and interdisciplinary research are needed. If we make the right choices now, we can enjoy the many benefits of AI, while minimising the risks of unfair discrimination.